\documentclass[12pt, letterpaper]{article}

\usepackage[ngerman,british]{babel}
\usepackage{graphicx, caption}
\usepackage{epstopdf}
\usepackage{amsmath}
\usepackage{amssymb}
\usepackage{braket}
\usepackage{xcolor}
\usepackage[utf8]{inputenc}
\usepackage[margin=10pt,font=small,labelfont=bf,format=plain]{caption}
\usepackage[section]{placeins}
\usepackage{lmodern}
\usepackage[T1]{fontenc}
\usepackage[onehalfspacing]{setspace}

\newcommand{\abs}[1]{\lvert#1\rvert}

\begin{document}
\begin{Large}
\begin{onehalfspace}
\centering
\textbf{A VERSATILE QUANTUM SIMULATOR FOR COUPLED OSCILLATORS USING A 1D CHAIN OF ATOMS TRAPPED NEAR AN OPTICAL NANOFIBER}
\end{onehalfspace}
\end{Large}
\vspace{0.5 cm}
\begin{center}
   Daniela Holzmann $^{1,\dagger}$, Matthias Sonnleitner $^{1}$ and Helmut Ritsch $^{1}$\\
   \vspace{0.5 cm}
$^{1}$  Institute for Theoretical Physics, University of Innsbruck, Technikerstra\ss{}e 25, A-6020 Innsbruck, Austria\\
\vspace{0.5 cm}
\begin{footnotesize}
$~^\dagger$ Daniela.Holzmann@uibk.ac.at
\end{footnotesize}
\end{center}

\section{Abstract}
 The transversely confined propagating light modes of a nano-photonic optical waveguide or nanofiber can mediate effectively infinite-range forces. We show that for a linear chain of particles trapped within the waveguide's evanescent field, transverse illumination with a suitable set of laser frequencies should allow the implementation of a coupled-oscillator quantum simulator with time-dependent and widely controllable all-to-all interactions. At the example of the energy spectrum of oscillators with simulated Coulomb interactions we show that different effective coupling geometries can be emulated with high precision by proper choice of laser illumination conditions. Similarly, basic quantum gates can be selectively implemented between arbitrarily chosen pairs of oscillators in the energy basis as well as in a coherent-state basis. Key properties of the system dynamics and states can be monitored continuously by analysis of the out-coupled fiber fields.

\section{Introduction}
Suitably designed laser fields allow one to trap individual quantum particles at well defined locations and cool them to their motional ground state~\cite{kaufman2012cooling,sheremet2021waveguide}. Already some time ago it has been demonstrated  that trapping and cooling is also possible close to optical nano-structures and, in particular, in the vicinity of a tapered optical nanofiber~\cite{vetsch2010optical,goban2012demonstration,beguin2018observation}. Once trapped, the atoms interact with the evanescent field of light modes propagating within the fiber~\cite{beguin2018observation} exchanging energy and momentum. Thus the light strongly influences the atomic motion in the trap which in turn modifies the light propagation~\cite{markussen2020measurement,jones2020collectively,shomroni2014all,pivovarov2021single}. As photons within the fiber propagate over practically infinite distances they collectively couple to all atoms, which induces all-to-all long-range interactions~\cite{holzmann2016tailored}. In this way thousands of atoms can be trapped, which leads to strong collective effects~\cite{prasad2020correlating}.

The individual atom-atom coupling via resonant photon emission by one atom followed by absorption by a second atom is typically rather small~\cite{cirac2012atomic,chang2012cavity}, but it can already lead to spatial self-ordering of the atoms~\cite{Metzger2006fiber}. The induced force can be significantly increased if the atoms are transversely illuminated far off any internal atomic resonance to induce collective coherent scattering into the fiber~\cite{chang2013self,buonaiuto2021dynamical,griesser2013light}. Here the interference between the mode amplitudes created by scattering from different particles leads to gradient or dipole forces, which appear without changing the internal atomic state from spontaneous emission~\cite{holzmann2014self}. Properties of these forces can be modified by help of two laser frequencies~\cite{ostermann2014scattering}.

The interactions between the particles depend on the properties of the incoming light field. With careful choice of laser frequencies and powers almost arbitrary shapes of interaction forces can be synthesized~\cite{holzmann2018synthesizing}. In this work we give examples how this property could be used for quantum simulation~\cite{georgescu2014quantum,kim2021quantum,feynman1982simulating,hartmann2016quantum,longhi2011optical,noh2016quantum,tashima2019direct,huo2012quantum,angelakis2013mimicking} as well as for quantum computation. By designing the incoming light field we show, for example, how the interaction between ions can be simulated, even if they are ordered in 2D or 3D geometries. In contrast to quantum simulation with ions~\cite{davoudi2020towards,cirac1995quantum} we can even turn off the interactions between arbitrary pairs of particles. In the second part we describe the oscillator states as qubits and use this approach to design quantum gates~\cite{kewes2016realistic,paulisch2016universal} or produce entangled states~\cite{leong2020large,li2012robust,gonzalez2011entanglement}.

\section{Materials and Methods}
\subsection{Tailored coupling of the quantized motion of a trapped atom chain}

\begin{figure}[h]
\includegraphics[width=\textwidth]{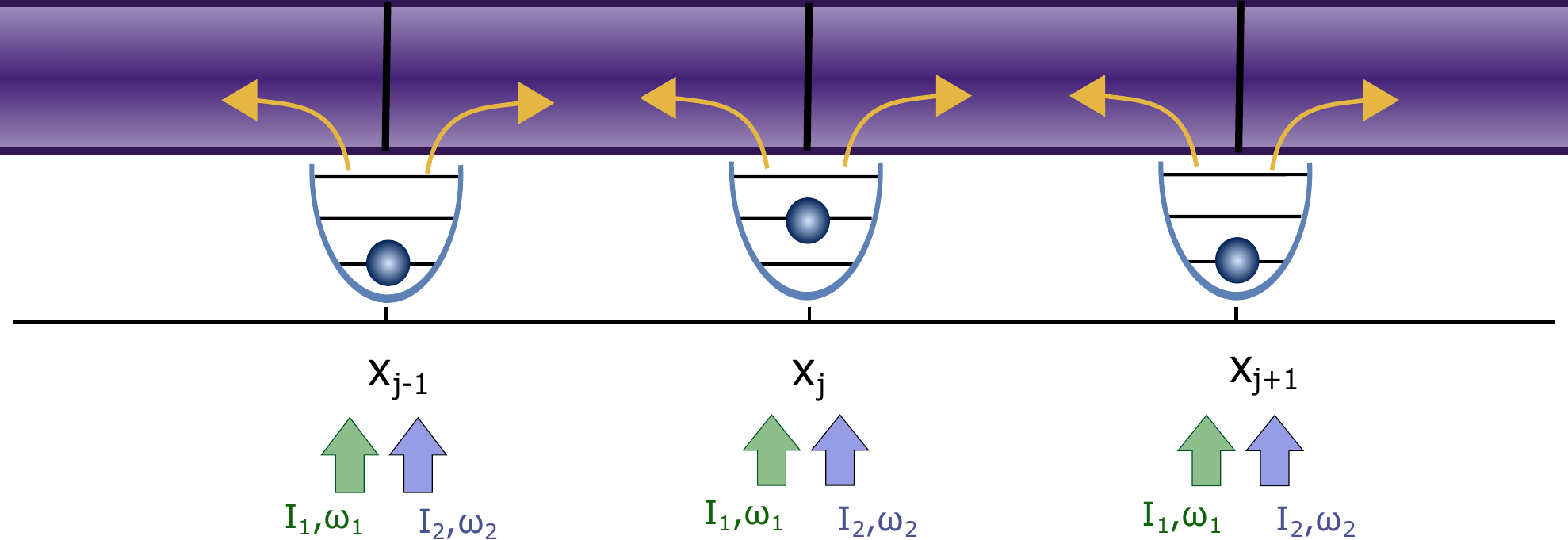}
\caption{Sketch of our system: $N$ particles are confined in homogeneous traps next to a nanofiber. The particles are illuminated by multi-color transverse pump fields and scatter light into the fiber. Interference of the scattered fields in the fiber leads to effective forces between the particles.}
\label{system}
\end{figure}

In this work we consider~$N$ particles, typically atoms, molecules or nanospheres, harmonically trapped at predefined positions along an optical nanofiber. As depicted in Figure~\ref{system}, these atoms interact with the evanescent field of the propagating nanofiber modes. Additionally, the particles are transversely illuminated by pump fields of tunable frequency. Each particle thus coherently scatters light from the pump fields into the fiber, where it interferes with light scattered by other atoms. The particles thus redistribute the field along the fiber which leads to effective interactions and forces between the particles. The interaction created by each frequency component of the pump light is long-range and depends on the distances between pairs of particles on the wavelength scale. Hence displacing one particle changes the overall fiber field and thus the forces acting on all other particles.

Similar long-range interactions and forces have been discussed already in a pioneering work by Chang et al. in~\cite{chang2013self}. There the focus was on resonant excitation and radiation pressure induced by internal transitions of an atom coupled to the waveguide. In line with our previous work~\cite{holzmann2018synthesizing} we will here allow for a very general form of mechanical interaction between the particles which can be achieved via frequency shaping of the illumination light.

In our model the transverse pump field is a sum of many plane waves with different intensities and frequencies. We assume that the different spectral components are sufficiently distinct such that the interference terms are negligible as the individual components are spatially but not time-coherent and interactions average out in time. The effective pair forces between the particles in such a system can be calculated using a beamsplitter matrix model describing all the scattering processes by the particles~\cite{holzmann2014self}.

Note that the particles in principle also back-scatter a fraction of the field propagating in the fiber. But since this contribution is typically very small and we assume weak coupling of the particles to the fiber field~\cite{chang2013self,holzmann2018synthesizing} such that each particle reflects only a tiny fraction of the propagating fiber fields. Hence we neglect back-scattering effects and assume that the force on the particles arises solely due to interference effects of the fields scattered into the fiber from the transverse pump. This assumption generally works well for small particle numbers, but for large system sizes even a small reflection by each particle can lead to significant collective effects~\cite{sheremet2021waveguide}.

Within this approximation the radiation force $F_{j}$ on a classical particle at position $x_j$ can be written as a sum of effective two-particle forces $f_\text{pair}(x_i,x_j)$ between this particle and all the other particles at positions $x_i$~\cite{holzmann2018synthesizing}

\begin{equation}\label{force_fj}
F_{j}=\sum_{\substack{i=1\\i \neq j}}^N f_\text{pair}(x_i,x_j)=\sum_{\substack{i=1\\i \neq j}}^N\sum_{k}\frac{\sigma_\text{sc} I_k \cos \left(k\left( x_j-x_i\right)\right)\text{sign}(x_i-x_j)}{c},
\end{equation}
with $N$ the number of particles along the fiber, $I_k$ the intensity of the field with frequency $\omega_k=k/c$ and $\sigma_\text{sc}$ the scattering cross section between the particles and the beam.

Using this force we define a two-particle potential $u_\text{pair}(x_i,x_j)$ such that $f_\text{pair}(x_i,x_j)=-\partial_{x_j} u_\text{pair}(x_i,x_j)$. For a system of $N$ particles the total potential is thus the sum of all two-particle interactions
\begin{equation}
 U_\text{tot}(x_1,\dots,x_N)=\frac{1}{2}\sum_{\substack{j=1}}^N \sum_{\substack{i=1 \\ i\neq j}}^N u_\text{pair}(x_i,x_j)=\frac{1}{2}\sum_{\substack{j=1}}^N \sum_{\substack{i=1 \\ i\neq j}}^N \sum_{k} \frac{\sigma_\text{sc}I_k}{ck}\sin\left(k\vert x_i-x_j\vert\right).
 \label{Utot}
\end{equation}

It is clear that any translation of one particle changes the light field along the fiber and thus the optical potential seen by all the other particles. The same result can be obtained following the model of Chang et al.~\cite{chang2013self} by taking the weak scattering limit for far-detuned light. Eliminating the internal excited states of the particles then leads to the force given in Equation~\eqref{force_fj}.

We studied such a system in a previous work~\cite{holzmann2018synthesizing} where we assumed classical point particles allowed to move freely along the fiber direction. Here we study locally trapped and very cold particles which requires a quantized description of motional degrees of freedom.

In the present model we thus consider particles trapped in harmonic potentials centered at positions $x_{j,0}$, $j=1,\dots, N$,  $U_\text{HO}=\sum_{i=1}^Nm\omega_{T}\Delta_i^2/2$, with $m$ the mass of the particles, $\omega_T$ the frequency of the harmonic oscillator traps and $\Delta_i = x_i-x_{i,0}$. We assume that the particles are tightly trapped in transverse direction and linearize the motion along the longitudinal motion of the particles in Equation~\eqref{Utot} around the center of the harmonic oscillators $x_{i,0}$, $x_i\rightarrow x_{i,0}+\Delta_i$, with $k \Delta_i\ll 1$. We thus obtain an effective Hamiltonian
\begin{multline}
\hat{H}=\sum_k\frac{I_k\sigma_\text{sc}}{c}\sum_{j=1}^N\left(\sum_{i=1}^{j-1}\left(\frac{1}{k}\sin\left(kd_{ij}\right)+\left(\Delta_j-\Delta_i\right)\cos\left(kd_{ij}\right)\right)\right.\\
\left.-\sum_{i=1}^{N}\frac{ k\left(\Delta_j-\Delta_i\right)^2}{4}\sin\left(kd_{ij}\right)\right)+\sum_{i=1}^N\left(\frac{P_i^2}{2m}+\frac{m\omega_{T}^2}{2}\Delta_i^2\right),
\end{multline}
with $d_{ij}=\vert x_{j,0}-x_{i,0}\vert$. We quantize the relative motion of the particles with respect to the trap centers by setting
\begin{subequations}
\begin{align}
\hat{\Delta}_i&=\sqrt{\frac{\hbar}{2m\omega_{T}}}(\hat{a}_i+\hat{a}_i^\dagger)=\delta_0 (\hat{a}_i+\hat{a}_i^\dagger),\\
\hat{P}_i&=i\sqrt{\frac{\hbar m\omega_{T}}{2}}(\hat{a}_i^\dagger-\hat{a}_i) = \frac{i\hbar}{2\delta_0} (\hat{a}_i^\dagger-\hat{a}_i)  ,
\end{align}
\end{subequations}
with the oscillator length $\delta_0^2=\hbar/(2 m \omega_T)$.

Ignoring the constant terms in the Hamiltonian we thus obtain
\begin{subequations}
\begin{align}
\hat{H}&=\hat{\tilde{H}}_\text{osc}+\hat{H}_\text{int}+\hat{H}_\text{rwa} \label{H}\\
\hat{\tilde{H}}_\text{osc}&=\sum_{i=1}^N\hbar \left(\omega_T-\sum_{j=1}^{N}\sum_{k}\Omega_k\sin\left(kd_{ij}\right)\right) \hat{a}_i^\dagger \hat{a}_i=\sum_{i=1}^N \hbar\tilde{\omega}_i\hat{a}_i^\dagger \hat{a}_i\\
\hat{H}_\text{int}&=\sum_{j=1}^N\sum_{i=1}^{N}\sum_{k}\hbar\Omega_k\sin\left(kd_{ij}\right)  \hat{a}_j^\dagger \hat{a}_i\\
\hat{H}_\text{rwa}&=\hbar\sum_{j=1}^N\left(\sum_{i=1}^{j-1}\sum_{k} \epsilon_k\cos\left(kd_{ij}\right)\left(\hat{a}_j+\hat{a}_j^\dagger-\hat{a}_i-\hat{a}_i^\dagger\right)\right.\notag\\
&\left.-\sum_{i=1}^{N}\sum_{k}\frac{\Omega_k}{2}\sin\left(kd_{ij}\right)\left(\hat{a}_j^2+\hat{a}_j^\dagger{}^2-\hat{a}_j \hat{a}_i-\hat{a}_j^\dagger \hat{a}_i^\dagger\right)\right),
\end{align}
\end{subequations}
with $\Omega_k:=\sigma_\text{sc}I_k\delta_0^2 k/(\hbar c)$ and $\epsilon_k:=\sigma_\text{sc}I_k\delta_0/(\hbar c)$.
Here, $\hat{H}_\text{rwa}$ generates force terms in the time evolution oscillating at the trapping frequency or higher, which typically cancel out when averaged over one period. Hence we will neglect them later in Section~\ref{quantumgates}. Effectively they lead to a small displacement of the particles equilibrium and a change of the effective frequency of the oscillators (squeezing). $\hat{H}_\text{int}$ describes the interactions between the particles and $\hat{\tilde{H}}_\text{osc}$ the harmonic potential with shifted frequency $\tilde{\omega}_i$ due to the interaction of the particles.

\subsection{Model assumptions and limitations }

In our model we assume that the particles are harmonically trapped and we require that the transverse pump fields convey strong enough particle-particle interactions to influence the motion of the trapped particles along the fiber direction. At the same time the fields must be weak enough such that we can  neglect saturation effects and eliminate any particle's internal degrees of freedom. In the following we study in some more detail how all these limiting conditions restrict the operating parameters.

The intensity of the light field scattered into the fiber by the particles depends on the incoming photon energy $\hbar\omega$ and flux, the emission rate into the fiber $\gamma_\text{guid}$, the effective mode cross section $A$ of the fiber field and the excited state population $\rho_{ee}$. We approximate the particles as effective two-level systems and operate at low saturation by choosing large laser detuning $\Delta\gg\Gamma$ from atomic resonance, with $\Gamma$ the decay rate. In this limit the excited state fraction is $\rho_{ee}\approx  I_0/(2I_\text{sat}({1+4\Delta^2/\Gamma^2})) \ll 1$, with $I_0$ the intensity of the incoming pump field and $I_\text{sat}$ the saturation intensity.

The emission rate into the fiber depends on the spatial profile of the fiber field determining the field strength at the atomic position and on the atomic dipole matrix element. A single-mode fiber carries only the fundamental $\mathrm{HE}_{11}$ -mode. The explicit expression for the mode profiles are, for example, given in~\cite{snyder1983optical}. Here, we assume that the modes and the atomic dipoles are linear polarized perpendicular to the fiber axis. In this case the particles scatter the light symmetrically into the fiber. Using Fermi's golden rule the emission rate into the fiber can be found with $\gamma_\text{guid} \approx 0.13~\Gamma$, with $\Gamma=\vert \vec{d}\vert^2\omega^3/(3\pi\epsilon_0\hbar c^3)$ the free-space emission rate~\cite{le2005spontaneous}. Usually $\gamma_\text{guid}$ for atoms along a nanofiber is between $0.1-0.2~\Gamma$~\cite{le2005spontaneous}, but can be tuned up to $0.99~\Gamma$ for quantum dots~\cite{scarpelli201999,liu2018high} or superconductory transmon qubits~\cite{mirhosseini2019cavity}. Recently the coupling efficiency could be improved for atoms, when using a hole-tailored nanofiber and reached $0.6~\Gamma$~\cite{wang2021high}.

The intensity scattered into the fiber by the particles can thus be estimated by
\begin{equation}
I_k=\frac{\hbar\omega}{A}\gamma_\text{guid}\rho_\text{ee}\approx \frac{1}{2} \frac{\hbar\omega}{A}\gamma_\text{guid}\frac{I_0/I_\text{sat}}{1+4\Delta^2/\Gamma^2}.
\label{intensity}
\end{equation}
For a nanofiber with radius $r=200~\mathrm{nm}$, the fiber cross section area $A=r^2\pi$, the Cesium D2-line $\omega\approx 2.2\cdot 10^{15}~\mathrm{Hz}$ with $\Gamma=33\cdot 10^6~\mathrm{1/s}$ and the mass of Cesium $m\approx220\cdot 10^{-27}~\mathrm{kg}$, and a detuning $\Delta = 100~\Gamma$ we find that the intensity scattered into the fiber is $I_k\approx 6.2\cdot10^{-4}~I_0/I_\text{sat}~\mathrm{W/m^2} $.

To linearize the interaction potential we require a very deep potential such that the particles are well trapped $ k \Delta_i \ll 1$. With $\Delta_i=\delta_0(a_i^\dagger+a_i)$ and $\delta_0=\sqrt{\hbar/(2m\omega_T)}$, this means

\begin{equation}
k \delta_0=\frac{\omega}{c}\sqrt{\frac{\hbar}{2m\omega_T}}\ll 1
\end{equation}
and we find
\begin{equation}\label{omega_trap}
\omega_T\gg \frac{\omega^2}{c^2}\frac{\hbar}{2m}.
\end{equation}
Using the parameters for the Cesium D2-line this implies $\omega_T\gg 10^5~\mathrm{Hz}$.

Using these requirements we can compare the trapping potential $H_T$ with the interaction arising from the transverse pump $H_\text{pump}$. Starting from the initial potential $\hat{H}=\hat{H}_\text{pump}+\hat{H}_T$ without any expansion

\begin{equation}
\hat{H}=\frac{1}{2}\sum_{j=1}^N\sum_{l=1}^{N}\frac{\sigma_\text{sc}I_k}{ck}\sin(k d_{jl})+\sum_{j=1}^N\hbar\omega_T \hat{a}_j^\dagger \hat{a}_j,
\end{equation}
we find $\hat{H}_\text{pump}\propto \sigma_\text{sc} I_k/(2k c)$ and $\hat{H}_T\propto \hbar\omega_T$. The scattering cross section can be approximated by the fiber cross section $\sigma_\text{sc}\approx A$. Inserting the intensity from Equation~\eqref{intensity} and the boundary on the trapping frequency from Equation~\eqref{omega_trap},  
\begin{align}
\frac{\hbar\omega_T}{\frac{\sigma_\text{sc}I_k}{2k c}}
\approx \frac{4\omega_T}{\gamma_\text{guid}\rho_\text{ee}}
= \frac{8\omega_T}{\gamma_\text{guid}}\frac{1+4\Delta^2/\Gamma^2}{I/I_\text{sat}}
\gg \frac{8}{\gamma_\text{guid}}\left(\frac{\omega}{c}\right)^2 \frac{\hbar}{2m}\frac{1+4\Delta^2/\Gamma^2}{I/I_\text{sat}}.
\end{align}
Using again the parameters for Cesium and setting $\rho_\text{ee}\approx 10^{-2}$ and $\omega_T\approx 10^6 Hz$ we get the condition $\hbar\omega_T/\left(\sigma_\text{sc}I_k/(2k c)\right) \gg 3$. 

Assuming tightly trapped particles the interaction strength in the Hamiltonian~\eqref{H} is thus
limited by
\begin{equation}
\Omega_k=\frac{\sigma_\text{sc}I_k\delta_0^2\omega}{\hbar c^2}\ll \frac{\gamma_\text{guid} I/I_\text{sat}}{2(1+\Delta^2/\Gamma^2)}.
\end{equation}
Using the parameters for Cesium as given above we find $\Omega_k\ll 10^5~\mathrm{Hz}$.

\section{Results}
The Hamiltonian in Equation~\eqref{H} shows that we can design the effective interactions between the particles by choosing the intensities and frequencies of the incoming fields as well as the distances between the trapping positions of the particles. In the following section we show how this can be used to simulate a system where particles interact via some specific physical potential of choice. Here as a generic long range interaction we choose a Coulomb-type $1/r$ potential. In the later Sections~\ref{quantumgates} and~\ref{entanglement} we discuss how this approach can be used to design quantum gates or how to entangle the motion of many particles.

\subsection{Simulating Coulomb interactions between trapped quantum particles}

The Hamiltonian in Equation~\eqref{H} can be used to simulate any symmetric two-body interaction. In the following example we will use the effective atom-atom interaction via the waveguide to mimic the Coulomb interaction between ions.

In principle one could tune the light fields to mimic a full Coulomb potential, but since $~1/r$ is difficult to approximate in a Fourier series this would require a very large number of laser fields. However, if we assume that the ions are also harmonically trapped, we only have to tune the atom-light interaction to mimic the Coulomb interaction at the position of the trapped ions.

To simulate the Coulomb (or any other) potential we first linearize it around the trapping positions, quantize the motional degrees of freedom and then compare the terms with the Hamiltonian from Equation~\eqref{H}. Using this concept one can even map higher-dimensional systems of interacting particles to our 1D-system.

\begin{figure}
\centering
    \includegraphics[width=0.9\textwidth]{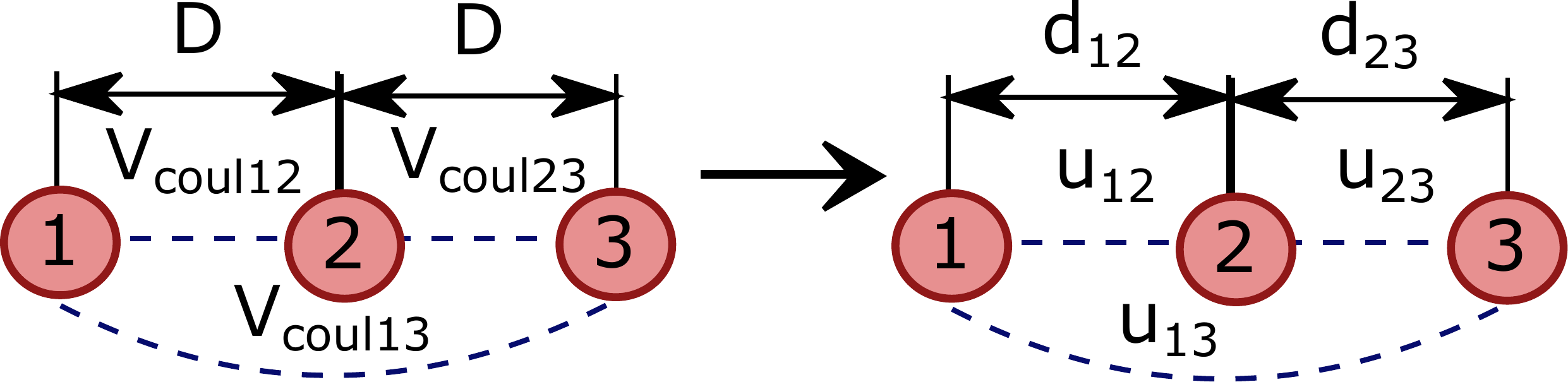}
\vspace{0.1\linewidth}

  \begin{minipage}[c]{0.48\linewidth}
    \includegraphics[width=\textwidth]{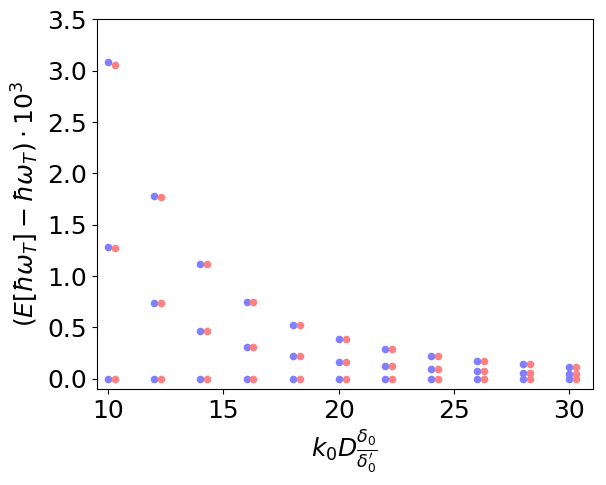}  
  \end{minipage} 
  \begin{minipage}[c]{0.48\linewidth}
    \includegraphics[width=0.96\textwidth]{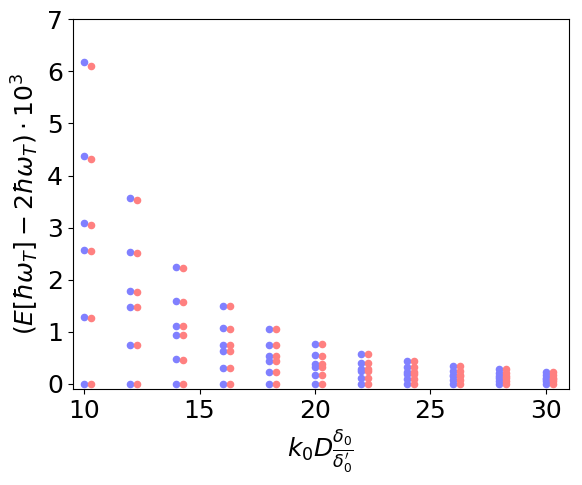} 
  \end{minipage}
  \hfill
  \caption{We map three ions along a line, which interact via the Coulomb force separated at a distance $D$ on a system of particles along a nanofiber, with distances $d_{12}$ and $d_{23}$. The lower figures show the eigenenergies of these three harmonically trapped ions with Coulomb repulsion (blue) in comparison to particles trapped along the fiber with simulated interaction (red) as function of the distance between the interacting particles. In the simulation the particles are trapped along the fiber at fixed distances $d_{12}=d_{23}=3/8~\lambda_0$ and the pump laser parameters are adjusted to mimic Coulomb interaction at arbitrary distances between the ions. We use $\Delta_k=0.7~k_0$ and $\Omega_0/\widetilde{\Omega}=0.34 + 1.07~k_0D\delta_0\delta_0^\prime/ $, $\Omega_1/\widetilde{\Omega}=1.16 + 1.34~k_0D\delta_0\delta_0^\prime/ $, $\Omega_2/\widetilde{\Omega}=1.68 + 1.58 ~k_0D\delta_0/\delta_0^\prime $, $\Omega_3/\widetilde{\Omega}=0.74+1.4~k_0D\delta_0/\delta_0^\prime$, with $\widetilde{\Omega}/\omega_T=\sum_l \Omega_l/\widetilde{\Omega}\left(10/k_0\delta_0^\prime/\delta_0\right)^3/D_{ij}^3/(0.004)$. The figure on the left side shows the energies corresponding to the first oscillator state and the right figure shows the energies corresponding to the second oscillator state.}
\label{eigenline} 
\end{figure}

Figure~\ref{eigenline} shows an example, where we simulate the interaction between three ions distributed along a line by a system of particles along a nanofiber. The Coulomb potential $V_\text{coul}$ of $N$ interacting ions along a line of charge $q_i$ at distances $D_{ij}$ is given by
\begin{equation}
     V_\text{coul}=\sum_{i=1}^N\sum_{j\neq i}\frac{1}{8\pi\epsilon_0} \frac{q_iq_j}{D_{ij}}.
\end{equation}
Linearizing around their trapping position, quantizing the motional degrees of freedom and ignoring the constant terms we find for $q=q_i=q_j$. 
\begin{subequations}
\begin{align}
\hat{H}_\text{coul}&=\hat{\tilde{H}}_{\text{coul}_\text{osc}}+\hat{H}_{\text{coul}_\text{int}}+\hat{H}_{\text{coul}_\text{rwa}}, \label{Hcoul}\\
\hat{\tilde{H}}_{\text{coul}_\text{osc}}&=\sum_{i=1}^N\hbar \left(\omega_T^\prime+\frac{1}{8\pi\epsilon_0\hbar}\sum_{j\neq i}\frac{4q^2\delta_0^{\prime 2} }{D_{ij}^3}\right) \hat{a}_i^\dagger \hat{a}_i,\\
\hat{H}_{\text{coul}_\text{int}}&=-\frac{1}{8\pi\epsilon_0}\sum_{i=1}^N\sum_{j\neq i}\frac{4q^2\delta_0^{\prime 2} }{D_{ij}^3} \hat{a}_i^\dagger \hat{a}_j,\\
\hat{H}_{\text{coul}_\text{rwa}}&=\frac{1}{8\pi\epsilon_0}\sum_{j=1}^N\left(\sum_{i=1}^{j-1}\frac{-2q^2\delta_0^\prime}{D_{ij}^2}\left(\hat{a}_j+\hat{a}_j^\dagger-\hat{a}_i-\hat{a}_i^\dagger\right)\right.\notag\\
&\left.+\sum_{j\neq i}\frac{2\delta_0^{\prime 2}q^2}{D_{ij}^3}\left(\hat{a}_j^2+\hat{a}_j^\dagger{}^2-\hat{a}_j \hat{a}_i-\hat{a}_j^\dagger \hat{a}_i^\dagger\right)\right)\label{coulrwa}.
\end{align}
\end{subequations}
This Hamiltonian from Equation~\eqref{Hcoul} describes tightly trapped ions interacting via a Coulomb potential.

To simulate this system with particles interacting via a waveguide we compare the terms of $H_\text{coul}$ with the Hamiltonian from Equation~\eqref{H}. We find that the individual terms agree if we choose distances~$d_{ij}$, wavenumbers~$k$ and interaction strengths~$\Omega_k$ such that:
\begin{subequations}\label{coulomb_equations}
\begin{align}
  \frac{\delta_0^\prime}{4\pi\epsilon_0\hbar\omega_T^\prime} \frac{q^2}{D_{ij}^2}=-\sum_k \frac{\epsilon_k}{\omega_T} \cos(k d_{ij}),\\
  \frac{\delta_0^{\prime 2}}{2\pi\epsilon_0 \hbar\omega_T^\prime} \frac{q^2}{D_{ij}^3}=-\sum_k \frac{\Omega_k}{\omega_T} \sin(k d_{ij}).
\end{align}
\end{subequations}
Note, that the distances between the particles in our system, $d_{ij}$, are different from the distances in the simulated Coulomb system, $D_{ij}$. Also the harmonic trapping frequencies $\omega_T$ and $\omega_T'$ can be chosen independently.

For given distances $d_{ij}$ and frequencies $k_l=k_0+l\Delta_k$, Equations~\eqref{coulomb_equations} become a set of linear equations for the interaction strengths $\Omega_l \equiv \Omega_{k_l}$. Choosing $D_{12}=D$ as a reference we define an interaction strength $\widetilde{\Omega}$ such that
\begin{equation}
\frac{q^2\delta_0^{\prime 2}}{2\pi\epsilon_0\hbar\omega_T^\prime}\frac{1}{D^3}=\frac{\widetilde{\Omega}}{\omega_T}.
\label{dist}
\end{equation}
Equations~\eqref{coulomb_equations} then simplify to
\begin{subequations}
\begin{align}
\frac{D^3}{2 D_{ij}^2}&=-\delta_0^\prime\sum_l\frac{\epsilon_l}{\widetilde{\Omega} } \cos((k_0+l\Delta_k) d_{ij})
\notag\\ &=
-\frac{\delta_0^\prime}{\delta_0}\sum_l\frac{\Omega_l}{\widetilde{\Omega} (k_0+l\Delta_k)} \cos((k_0+l\Delta_k) d_{ij}),\\
 \frac{D^3}{D_{ij}^3}&=-\sum_l \frac{\Omega_l}{\widetilde{\Omega}} \sin((k_0+l\Delta_k) d_{ij}).
\end{align}
 \label{couleq}
\end{subequations}
These equations form a linear system of equations for the interaction strength depending on the intensities of the incoming fields. Consequently, we have two equations for every different distance $D_{ij}$ and thus need the same number of fields. One way to solve this system is to assume that the particles are equally distributed with $d_{i,i+1}=3\lambda_0/8$ and then choose the frequency spacing $\Delta_k$ such that the intensities $I_k$ are positive.

Although the fiber system is a 1D-system, one could even map 2D or 3D-systems on it. In this case we have $N\cdot N_D$ oscillators, with $N$ the number of ions and $N_D$ the number of dimensions. And in general this means we also need $N\cdot N_D$ particles in the simulation. Here the first $N$ particles correspond to the interactions between the ions in the first dimension and the second $N$ particles to the interactions in the second dimension and so on.

\begin{figure}
\centering
    \includegraphics[width=\linewidth]{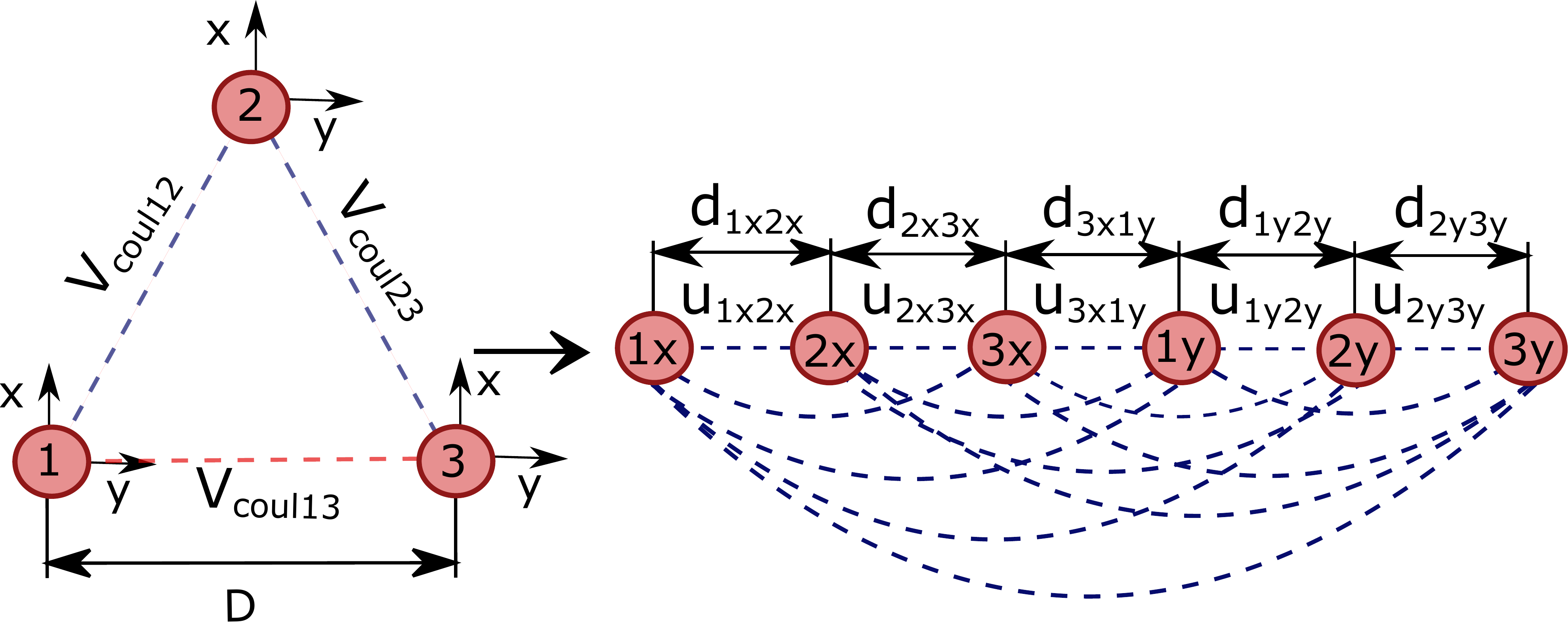}
  \caption{We simulate the Coulomb interaction of three ions arranged in an equilateral triangle by a 1-D particle chain along a nanofiber. The first three particles correspond to the interaction in x-direction, while the next three particles correspond to the interaction in y-direction. In Fig.~\ref{triangle1} we also show an example where the interaction between the ions number 1 and 3 is turned off.}
\label{triangle} 
\end{figure}

Figure~\ref{triangle} shows three ions arranged in an equilateral triangle. In this case we have $N_D=2$ dimensions and $N=3$ ions and thus need six particles along the fiber. The Coulomb potential then is given by
\begin{equation}
V_{\text{coul}}=\sum_{i=1}^N\sum_{i\neq j}\frac{1}{8\pi\epsilon_0}\frac{q_i q_j}{\sqrt{D_{ij_x}^2+D_{ij_y}^2}},
\end{equation}
with $D_{ij_x}$ and $D_{ij_y}$ the distances between the ions in $x$ and $y$-direction and $D_{ij}^2=D_{ij_x}^2+D_{ij_y}^2$. Here we have to linearize in the $x$ as well as in the $y$ direction. We assume that the oscillators in $x$ and $y$ directions have the same trapping frequency $\omega_T'$. To reduce the number of equations, we here ignore the fast oscillating terms $\hat{H}_{\text{coul}_\text{rwa}}$ and find
\begin{subequations}
\begin{align}
\hat{H}_\text{coul}&=\hat{\tilde{H}}_{\text{coul}_\text{osc}}+\hat{H}_{\text{coul}_\text{int}},\\
\hat{\tilde{H}}_{\text{coul}_\text{osc}}&=\sum_{i=1}^N\hbar \left(\omega_T^\prime+\frac{1}{8\pi\epsilon_0\hbar}\sum_{j\neq i}\frac{4q^2\delta_0^{\prime 2} }{D_{ij}^5}\right) \left(\left(D_{ij_x}^2-\frac{1}{2}D_{ij_y}^2\right)\hat{a}_{i_x}^\dagger \hat{a}_{i_x}\right.\nonumber\\
&\left.+\left(D_{ij_y}^2-\frac{1}{2}D_{ij_x}^2\right)\hat{a}_{i_y}^\dagger \hat{a}_{i_y}\right),\\
\hat{H}_{\text{coul}_\text{int}}&=-\frac{1}{8\pi\epsilon_0}\sum_{i=1}^N\sum_{j\neq i}\frac{4q^2\delta_0^{\prime 2} }{D_{ij}^5} \left( \left(D_{ij_x}^2-\frac{1}{2}D_{ij_y}^2\right) \hat{a}_{i_x}^\dagger \hat{a}_{j_x}+\left(D_{ij_y}^2-\frac{1}{2}D_{ij_x}^2\right) \hat{a}_{i_y}^\dagger \hat{a}_{j_y}\right.\nonumber\\
&\left.-\frac{3}{2}D_{ij_x}D_{ij_y} \left(\hat{a}_{i_x}^\dagger \hat{a}_{j_y}+\hat{a}_{i_x} \hat{a}_{j_y}^\dagger-\hat{a}_{i_x} \hat{a}_{i_y}^\dagger-\hat{a}_{i_x}^\dagger \hat{a}_{i_y}\right)\right).
\end{align}
\end{subequations}
Defining an interaction strength $\widetilde{\Omega}$, $D=D_{12}$ and $k_l=k_0+l\Delta_k$ as in Equation~\eqref{dist} we have to solve the following 15 equations
\begin{subequations}
\begin{align}
     \frac{D^3}{D_{ij}^5}\left(D_{ij_x}^2-\frac{1}{2}D_{ij_y}^2\right)&=-\sum_l \frac{\Omega_l}{\widetilde{\Omega}} \sin((k_0+l\Delta_k) d_{i_xj_x}),\\
     \frac{D^3}{D_{ij}^5}\left(D_{ij_y}^2-\frac{1}{2}D_{ij_x}^2\right)&=-\sum_l \frac{\Omega_l}{\widetilde{\Omega}} \sin((k_0+l\Delta_k) d_{i_yj_y}),\\
     \frac{3}{2}\frac{D^3}{D_{ij}^5}D_{ij_x}D_{ij_y}&=\sum_l \frac{\Omega_l}{\widetilde{\Omega}} \sin((k_0+l\Delta_k) d_{i_xj_y}),\\
     \sum_j\frac{3}{2}\frac{D^3}{D_{ij}^5}D_{ij_x}D_{ij_y}&=-\sum_l \frac{\Omega_l}{\widetilde{\Omega}} \sin((k_0+l\Delta_k) d_{i_xi_y}).
\end{align}
\label{couleqtriangle}
\end{subequations}

A special feature of such a mapping is that the interactions between specific pairs of particles can be individually tuned or even turned off as in Figure~\ref{triangle} for the particles at the bottom of the triangle. This allows one to implement any graph of interacting particles. Obviously such systems could not be implemented with actual ions.

\begin{figure} 
  \begin{minipage}[c]{0.48\linewidth}
    \includegraphics[width=\textwidth]{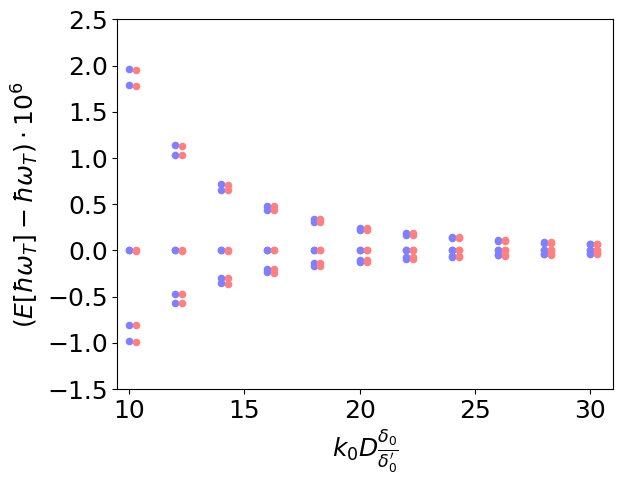}  
  \end{minipage} 
  \begin{minipage}[c]{0.48\linewidth}
    \includegraphics[width=0.96\textwidth]{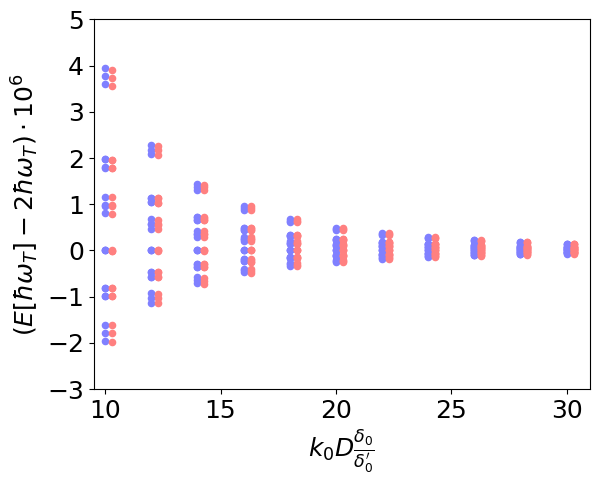} 
  \end{minipage}
  \hfill
   \begin{minipage}[c]{0.48\linewidth}
    \includegraphics[width=\textwidth]{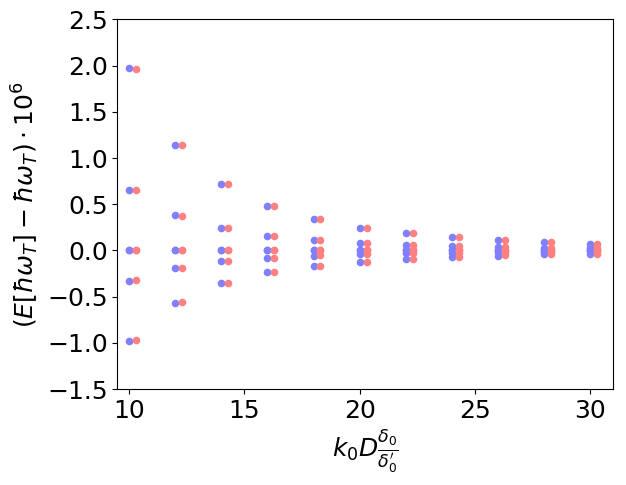}  
  \end{minipage} 
  \hfill
  \begin{minipage}[c]{0.48\linewidth}
    \includegraphics[width=0.96\textwidth]{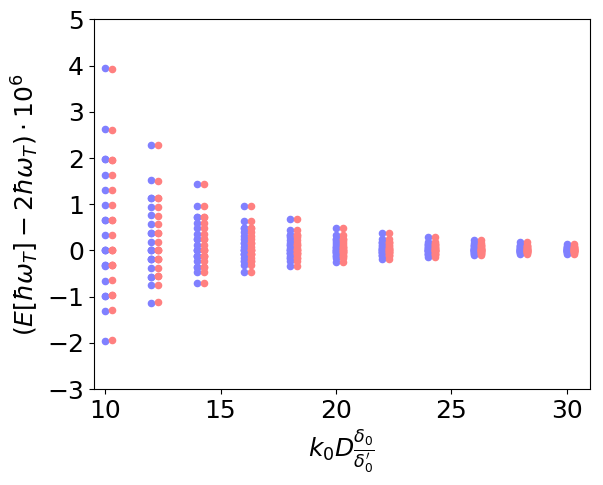} 
  \end{minipage}
  \caption{Eigenenergies of three harmonically trapped ions with Coulomb repulsion ordered like a triangle (blue) in comparison to particles trapped along a fiber with light-induced interaction (red) as shown in Figure~\ref{triangle} as function of the distance between the interacting ions. In the simulation the particles are trapped along the fiber at fixed distances $d_{12}=d_{23}=1/3~\lambda_0$, $d_{34}=\lambda_0$ and $d_{45}=d_{56}=1/4~\lambda_0$ and the pump laser parameters are adjusted to mimic Coulomb interaction at arbitrary distances between the ions with $\Delta_k=0.33~k_0$ and $\widetilde{\Omega}/\omega_T=\sum_l \Omega_l/\widetilde{\Omega}\left(10/k_0\delta_0^\prime/\delta_0\right)^3/D_{ij}^3/(0.004)$. The figures in the upper row show the eigenenergies when all particles are interacting, while in the lower figures the interaction between ions number 1 and 3 at the bottom of the triangle is suppressed. The figures on the left side show the energies corresponding to the first oscillator state and the right figures show the energies corresponding to the second oscillator state. Data values can be found in the appendix in Table~\ref{data}.}
\label{triangle1} 
\end{figure}

To show the validity of this mapping we calculate the eigenenergies of the waveguide system and compare them with the original Coulomb system (cf. Figure~\ref{eigenline} for the ions along a line and Figure~\ref{triangle1} for the ions ordered like a triangle). The energy levels of the oscillators split up due to the interaction between the particles. As long as the tight-binding condition for the ions is met, we find excellent agreement for all three systems. Note that the frequencies to solve Equations~\eqref{couleq} and~\eqref{couleqtriangle} are distributed over a large spectrum between $k_0$ and $3.8~ k_0$ and $k_0$ and $5.3~k_0$, respectively. But other methods to solve the equations might avoid this issue.

\subsection{Bipartite quantum gates between distant particles }\label{quantumgates}
In the previous section we showed how changing the distances between the traps or the intensity and frequency of the incoming light fields allows one to tailor the interaction between the particles. Here we demonstrate how this can be used to design quantum gates.

Writing the  Hamiltonian from Equation~\eqref{H} in an interaction picture we find that the terms $\propto \hat{a}_i,~\hat{a}_i^\dagger$ oscillate with $\tilde{\omega}_i$ and the terms $\propto \hat{a}_i^2,~\hat{a}_i^\dagger{}^2,~\hat{a}_j\hat{a}_i,~\hat{a}_j^\dagger \hat{a}_i^\dagger $ oscillate even with $2\tilde{\omega}_i$, while the terms $\propto \hat{a}_j^\dagger \hat{a}_i,~\hat{a}_j^\dagger \hat{a}_j $ do not oscillate. The rapidly oscillating terms average to zero and thus the Hamiltonian of Equation~\eqref{H} simplifies to

\begin{equation}
\hat{H}_{int}
=\sum_{j=1}^N \sum_{i=1}^{N}\sum_{k}\hbar\Omega_{k}\sin\left(k\vert x_{j,0}-x_{i,0}\vert\right)a_j^\dagger a_i
= \sum_{i,j=1}^N \hbar g_{ij} a_j^\dagger a_i
,
\label{Hrwa}
\end{equation}
with $\Omega_k=\sigma_{sc}I_k\delta_0^2/(\hbar k c)$ and $g_{ij}:= \sum_k \Omega_{k}\sin\left(k d_{ij}\right)$. Now it is obvious that the interaction between any particle pair $i$ and $j$ can be turned off by finding frequencies, positions and intensities such that the coupling~$g_{ij}$ vanishes. This is very important for gates as they should only act on special and not on all particles.

\begin{figure}
\center
\includegraphics[width=0.5\textwidth]{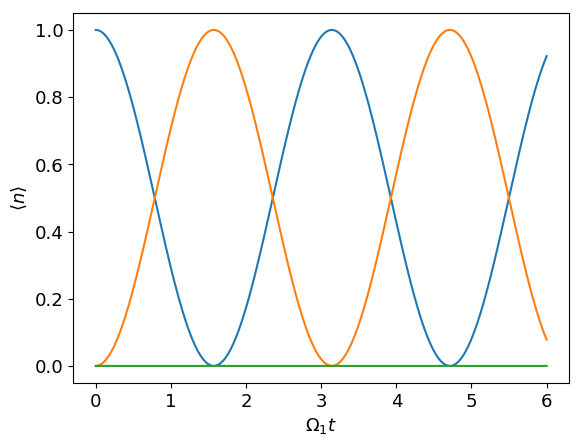}
\caption{Time evolution of the excited motional state occupation for three coupled particles. The blue line corresponds to the first particle, the orange line to the second particle and the green line to the third particle. We start from a state $\vert 100\rangle$ where only the first particle is excited and set the distances to $d_{12}=3/4~\lambda_1$, $d_{23}=7/8~\lambda_1$, $k_2=4/3 k_1$ and $\Omega_2=0.82~\Omega_1$. Choosing these parameters, the interaction between the third and the other two particles can be turned off, while the first two particles still interact with each other.}
\label{n12}
\end{figure}

Figure~\ref{n12} shows an example for three particles, where only two particles interact. Choosing the distances and frequencies with $k_l=k_0+l\Delta_k$ with $\Delta_k<k_0$, we have to solve $N(N-1)/2$ equations, with $N$ the number of the particles in the system and, thus, need $N(N-1)/2$ frequencies and intensities. Note, that the distances have to be chosen such that they are different for interacting and non-interacting pairs.

\subsubsection{Using the two lowest oscillator states as qubit basis}
The simplest way to map the harmonic oscillator to a qubit-system is to consider only the ground $\vert 0\rangle$ and first excited state $\vert 1\rangle$. As the Hamiltonian from Equation~\eqref{Hrwa} is $\propto a_j^\dagger a_i$ the resulting dynamics do not leave the subspace of these two states. For the basis ($\vert 0\rangle \vert 0\rangle$, $\vert 0\rangle \vert 1\rangle$, $\vert 1\rangle \vert 0\rangle$ and $\vert 1\rangle \vert 1\rangle$), the states evolve with the time-evolution operator $\hat{U}(t)= \exp(-i \hat{H} t/\hbar)$,

\begin{equation}
    \hat{U}(t)=\begin{pmatrix}
1&0&0&0\\
0&\cos(2 gt)&-i\sin(2 gt)&0\\
0&-i\sin(2gt)&\cos(2gt)&0\\
0&0&0&1
\end{pmatrix},
\end{equation}
which is equivalent to a mapping

\begin{subequations}\label{mapping_fockstates}
\begin{align}
    \vert 0\rangle \vert 0\rangle &\rightarrow \vert 0\rangle \vert 0\rangle\\
    \vert 0\rangle \vert 1\rangle&\rightarrow \cos(2 gt) \vert 0\rangle\vert 1\rangle-i\sin(2 gt)\vert 1\rangle \vert 0\rangle\\
    \vert 1\rangle \vert 0\rangle&\rightarrow -i\sin(2 gt)\vert 0\rangle \vert 1\rangle+\cos(2 gt) \vert 1\rangle \vert 0\rangle\\
    \vert 1\rangle \vert 1\rangle&\rightarrow \vert 1\rangle \vert 1\rangle
\end{align}
\end{subequations}
In this case the states $\vert 0\rangle \vert 0\rangle$ and $\vert 1\rangle \vert 1\rangle$ are not affected by the interaction, but we see oscillations between $\vert 0\rangle \vert 1\rangle$ and $\vert 1\rangle \vert 0\rangle$.

After an interaction time such that $gt=\pi/4+2\pi n$, $n\in\mathbb{Z}$,  $U(t)$ changes to
\begin{equation}
\hat{U}_\text{SWAP}=\begin{pmatrix}
1&0&0&0\\
0&0&-i&0\\
0&-i&0&0\\
0&0&0&1
\end{pmatrix}.
\end{equation}
This corresponds to an i-SWAP gate, which swaps the states of the two particles and introduces a phase, if the two particles are in different states.

Similarly the square root of an i-SWAP-gate (SQiSW) can be implemented by choosing $g_{ij}t=\pi/8+\pi n$, $n\in\mathbb{Z}$. Then $U(t)$ changes to

\begin{equation}
\hat{U}_\text{SQiSW}=\begin{pmatrix}
1&0&0&0\\
0&\frac{1}{\sqrt{2}}&-\frac{i}{\sqrt{2}}&0\\
0&-\frac{i}{\sqrt{2}}&\frac{1}{\sqrt{2}}&0\\
0&0&0&1
\end{pmatrix}.
\label{U2}
\end{equation}
$\hat{U}_\text{SQiSW}$ is a universal entangling gate and any quantum computation can be implemented using only single qubit rotations and the SQiSW-gate~\cite{bogdanov2012quantum}. However note that single qubit rotations cannot be implemented in the formalism described here as every interaction changes the state of (at least) two particles. One would thus need a separate mechanism to  rotate the state of each particle individually.

\subsubsection{Coherent states as computational basis}
As our individual quantum systems are oscillators we can go beyond the two-state approximations and also use higher excited motional states as computational basis. One particularly useful approach, which has been put forward and intensively studied for photons, is the use of coherent states as qubits. Typically the computational basis is then a pair of coherent states $\vert -\alpha\rangle,~\vert +\alpha\rangle$ ~\cite{kok2007linear,jeong2002efficient,ralph2003quantum,marek2010elementary}, 
\begin{equation}
\vert \alpha\rangle=\sum_{i=0}^{\infty}e^{-\frac{\vert\alpha\vert^2}{2}}\frac{\alpha^i}{\sqrt{i!}}\vert i\rangle,
\end{equation}
with complex amplitude $\alpha$. 
Although the two states are not perfectly orthogonal, the overlap between $\vert +\alpha \rangle$ and $\vert -\alpha \rangle$ is negligibly small for sufficiently large $\abs{\alpha}$. For example,
\begin{equation}
     \vert\langle +\alpha \vert -\alpha \rangle\vert^2=e^{-4\vert\alpha\vert^2} \approx 0.018,
\end{equation}
for amplitudes as small as $\alpha=1$.
In this basis quantum calculations can be performed relatively loss and fault tolerant~\cite{ralph2003quantum} and it turns out that all relevant two qubit interactions can be based on the so-called beamsplitter coupling between two sites~\cite{marek2010elementary}. The interaction is then simply given by $U(t)=\exp\left(i\theta/2\left(\hat{a}_1^\dagger\hat{a}_2+\hat{a}_1\hat{a}_2^\dagger\right) \right)$, with $\theta$ the polarization angle between the two interacting beams. It turns out that this is just the dominant term of light scattering interaction~\eqref{Hrwa} which can be well controlled in strength, time and space. 

The subset of coherent states $ \{\vert \alpha\rangle\vert \alpha^\prime\rangle\}$, with $\abs{\alpha}=\abs{\alpha'}$ evolves as
\begin{align}
e^{i\hat{H}_{int}t/\hbar}\vert \alpha\rangle\vert\alpha^\prime\rangle= \vert\alpha\cos\left( gt\right)+i\alpha^\prime\sin\left( gt\right)\rangle\vert\alpha^\prime\cos\left( gt\right)+i\alpha\sin\left( gt\right)\rangle.
\end{align}
such that
\begin{subequations}
\begin{align}
    \vert-\alpha\rangle\vert-\alpha\rangle&\rightarrow \vert -e^{igt}\alpha\rangle\vert -e^{igt}\alpha \rangle\\
    \vert-\alpha\rangle\vert+\alpha\rangle&\rightarrow \vert -e^{-igt}\alpha\rangle\vert e^{-igt}\alpha \rangle\\
    \vert+\alpha\rangle\vert-\alpha\rangle&\rightarrow \vert e^{-igt}\alpha\rangle\vert -e^{-igt}\alpha \rangle\\
    \vert+\alpha\rangle\vert+\alpha\rangle&\rightarrow \vert e^{igt}\alpha\rangle\vert e^{igt}\alpha \rangle
\end{align}
\end{subequations}
A more detailed calculation of this evolution can be found in the Appendix~\ref{evolcoh}.
As discussed in~\cite{ralph2003quantum}, this evolution corresponds to a beamsplitter interaction for photonic states. There it is also discussed that one can use this and a single-qubit rotation to implement a CNOT gate.

In contrast to what we found in Equation~\eqref{mapping_fockstates}, we here see that the state $\vert -\alpha\rangle\vert-\alpha\rangle$ can be flipped to $\vert+\alpha\rangle\vert+\alpha\rangle$ and vice versa. Note also that the coherent qubits evolve outside the subspace~$\{ \vert +\alpha\rangle, \vert-\alpha\rangle \}$ for $g t \neq n \pi$.

\subsection{Entanglement propagation via controlled long-range interaction}
\label{entanglement}
The discussion above focused on entangling any two particles in a larger system using quantum gates realized by two-particle-gates. Here we shall briefly investigate how a larger number of particles can be entangled.

If we only have a single pump field of frequency $k_0$ and put all particles at equal distance $n \pi k_0$, with arbitrary integer $n$, then no particle will interact with any other particle as $\sin\left(k_0 d_{ij}\right)=0$. If we now displace one particle by $\zeta \neq n \pi k_0$, this particle starts to interact with all other particles, but there are still no direct interactions between the remaining particles. 

\begin{figure}
\center
\includegraphics[width=0.5\textwidth]{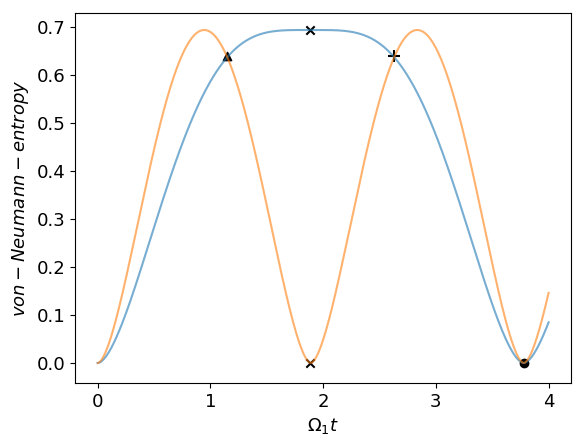}
\caption{Entanglement buildup for three coupled particles as a function of time with $d_{12}=\lambda_0/2$ and $d_{23}=\left(1/2+0.1\right)\lambda_0$ for the initial state $\vert 001\rangle$. The two curves show the entanglement entropy of the subsystem containing particles 2 and 3 (blue) and the subsystem containing particles 1 and 2 (yellow). So, the blue line describes the entanglement between the subsystem containing particle 1 and the subsystem containing particles 2 and 3, and the yellow line between the subsystem containing particle 3 and the subsystem containing particles 1 and 2. ($\blacktriangle$) corresponds to the state $1/\sqrt{3}\left(\vert 001\rangle-i\vert 010\rangle+i\vert 100\rangle\right)$, ($\times$) to the state $1/\sqrt{2}\left(\vert 01\rangle-\vert 10\rangle\right)\vert 0\rangle$ and ($+$) to the state $\frac{1}{\sqrt{3}}\left(-\vert 001\rangle+i\vert 010\rangle-i\vert 100\rangle\right)$ and (\textbullet ) to the state $-\vert 001\rangle$.}
\label{vne}
\end{figure}

But as shown in Figure~\ref{vne}, this is sufficient to create an effective all-to-all interaction. There we have three particles where the first and the second particle do not interact directly, but both interact with the third particle.

This is demonstrated using the mutual von-Neumann entropy,
\begin{equation}
    S_i=-\text{tr}\left(\rho_i\ln\rho_i\right),
\end{equation}
with $\rho_i$ being the reduced density matrix of the subsystem $i$.

In the left plot of Figure~\ref{vne} we start with a pure state, $\vert 001 \rangle$. But after a time such that $\cos\left(2\sqrt{2}\Omega_1 \sin\left(k\zeta\right)t\right) =1/\sqrt{3}$, indicated by the triangle, all three particles are entangled. Later, at the time indicated by the ($\times$), particle one and two are maximally entangled with each other, but disentangled from the third particle.

\subsection{State read out via the outgoing fiber fields}

In the previous chapters we discussed how the motional states of the particles can be manipulated, but how would such a manipulation be measured?

The fields leaving the fiber at the left and right edges contain information about the states of the particles in the system and by measuring the outgoing intensities one can determine the states of the particles.

Following the beamsplitter matrix formalism introduced in~\cite{holzmann2014self} we find for the amplitudes of the outgoing fields to the left $E_-(x_1)$ and to the right $E_+(x_N)$ of a system with $N$ particles
,
\begin{subequations}
\begin{align}
    E_-(x_1)&=\sum_k\sum_{i=1}^N \sqrt{\frac{I_k}{c\epsilon_0}} e^{ik(x_i-x_1)},\\
    E_+(x_N)&=\sum_k\sum_{i=1}^N \sqrt{\frac{I_k}{c\epsilon_0}} e^{ik(x_N-x_i)}.
\end{align}
\end{subequations}
As the particles are well trapped we can linearize these amplitudes as we did for the Hamiltonian and find

\begin{subequations}
\begin{align}
    \hat{E}_-(x_1)&=\sum_k\sum_{i=1}^N \sqrt{\frac{I_k}{c\epsilon_0}}\left( e^{ik(x_i-x_1)}+ik\delta_0 e^{ik(x_i-x_1)} \left(\hat{a}_i+\hat{a}_i^\dagger-\hat{a}_1-\hat{a}_1^\dagger\right)\right.\\
    \nonumber&\left.-\frac{1}{2}k^2\delta_0^2 e^{ik(x_i-x_1)} \left(\hat{a}_i+\hat{a}_i^\dagger-\hat{a}_1-\hat{a}_1^\dagger\right)^2\right),\\
    \hat{E}_+(x_N)&=\sum_k\sum_{i=1}^N \sqrt{\frac{I_k}{c\epsilon_0}}\left( e^{ik(x_N-x_i)}+ik\delta_0 e^{ik(x_N-x_i)} \left(\hat{a}_N+\hat{a}_N^\dagger-\hat{a}_i-\hat{a}_i^\dagger\right)\right.\\
    &\left.-\frac{1}{2}k^2\delta_0^2 e^{ik(x_N-x_i)} \left(\hat{a}_N+\hat{a}_N^\dagger-\hat{a}_i-\hat{a}_i^\dagger\right)^2\right).\nonumber
\end{align}
\end{subequations}
This way we can calculate the expectation vales for amplitudes and intensities for any given particle state.

\begin{figure}
\begin{minipage}[c]{\linewidth}
\centering
    \includegraphics[width=0.4\linewidth]{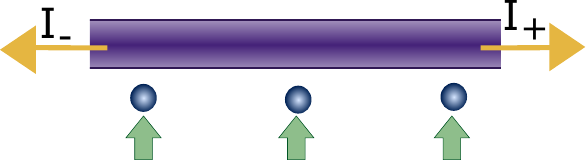}  
  \end{minipage} 
  
 \begin{minipage}[b]{0.5\linewidth}
    \includegraphics[width=\linewidth]{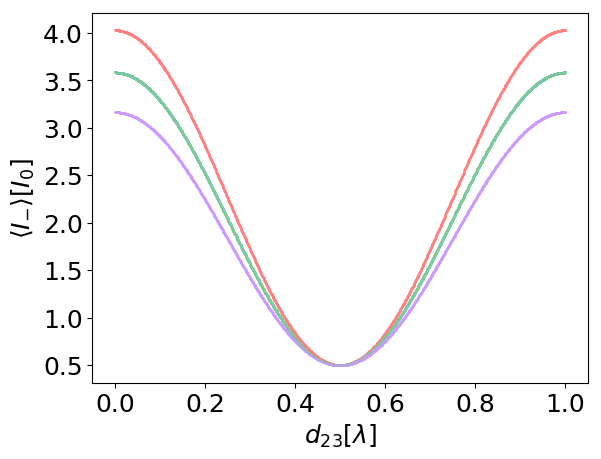}  
  \end{minipage} 
  \begin{minipage}[b]{0.5\linewidth}
    \includegraphics[width=\linewidth]{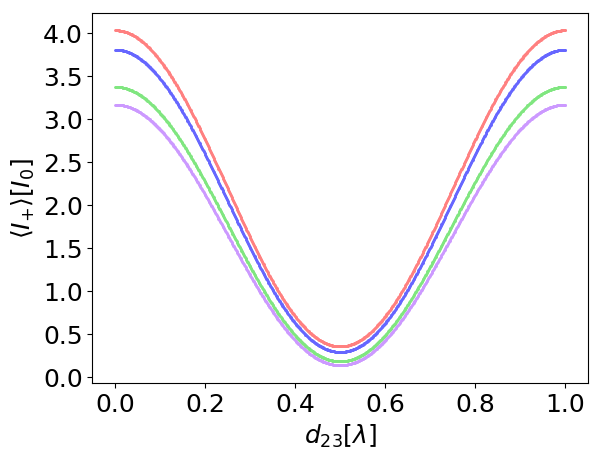} 
  \end{minipage}
  \caption{State-dependent light intensities emitted from the fiber to the left $I_-$ and to the right $I_+$ for a system with three particles. Here the distance between the first two particles stays constant with $d_{12}=\lambda_0$, while we vary the position of the third particle. Red lines correspond to the ground state $\vert 000\rangle$, blue lines to the a single excited state $\vert 100\rangle$, green lines to a doubly excited state $\vert 011\rangle$ and purple lines corresponds to the state $\vert 111\rangle$. Note, that in the left figure the green and blue line overlap.} 
  \label{I3} 
\end{figure}

\begin{figure}
 \center
    \includegraphics[width=0.5\textwidth]{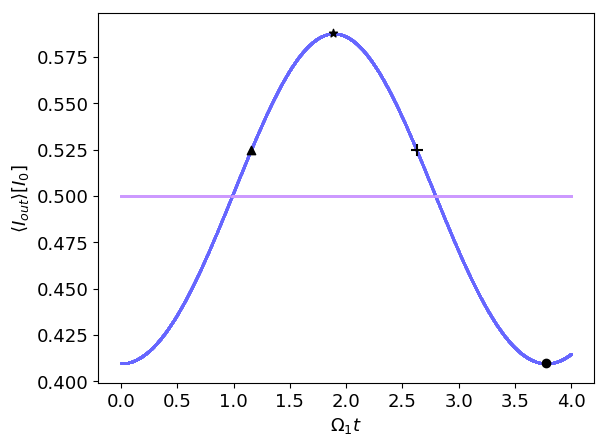}  
\caption{Average output power $I_-$ emitted on the left side of the fiber and $I_+$  on the right side for a system of two particles as a function of time. The initial condition is the same as in Figure~\ref{vne}. We start with the state $\vert 001\rangle$ and the particles are placed at a distance $d_{12}=1/2\lambda_0$ and $d_{23}=(\frac{1}{2}+0.1)\lambda_0$. The blue line corresponds to the outgoing intensity to the left side $\langle I_- \rangle$ and the purple line to the (constant) outgoing intensity to right $\langle I_+ \rangle$. ($\blacktriangle$) corresponds to the state $\frac{1}{\sqrt{3}}\left(\vert 001\rangle-i\vert 010\rangle+i\vert 100\rangle\right)$, ($\star$) to the state $\frac{1}{\sqrt{2}}\left(\vert 01\rangle-\vert 10\rangle\right\vert 0\rangle$, ($+$) to the state $\frac{1}{\sqrt{3}}\left(-\vert 001\rangle+i\vert 010\rangle-i\vert 100\rangle\right)$ and ($\cdot$) to $-\vert001\rangle$ }
\label{int001}
\end{figure}

Figure~\ref{I3} shows an example for the outgoing intensity expectation values for three particles. It confirms that the outgoing intensity depends on the states of the particles. Here, the states $\vert 100\rangle$ and $\vert 011\rangle$ cannot be distinguished in the left outgoing intensity, $I_-$, but they can be distinguished in $I_+$.

In Figure~\ref{int001} we plot the outgoing intensity for the initial conditions as above in Figure~\ref{vne}. From the outgoing intensity we can learn which particles are entangled and which are not.

\section{Discussion}

Mechanical interactions of particles trapped in the vicinity of an optical nanofiber can be controlled in a versatile form by choosing the properties of incoming transverse pump light. Using spatial and spectral light shaping of the illumination lasers the interactions between the particles can be tailored to simulate a wide class of interaction potentials between the particles. 

Here we studied the low temperature limit using a quantum mechanical description of the particle motion along the fiber direction at the trap sites and couple the particles via a nonlocal interaction through collective coherent light scattering into the fiber. We demonstrate that this system can be used to simulate, for example, Coulomb interactions between harmonically trapped particles with high precision. The idea can be extended in a straightforward manner beyond linear equidistant chains to effectively mimic a very general class of geometries including 2D-configurations. Using time dependent laser illumination one can even turn on and off specific interactions between arbitrary particle pairs simultaneously. By monitoring the spectrum and intensity of the light scattered out of the fiber ends ample information on the particle motion can be extracted in a minimally invasive way.

As another natural application the system offers variable possibilities to design two-qubit-gates, using not only oscillator eigenstates but also coherent states as computational basis. The virtually infinite range of the fiber mediated interaction should allow to implement larger systems of many qubits, without the requirement of closely spaced trapping sites allowing independent pairwise addressing control of quantum gates. As generic examples we studied the preparation of multi partite entangled states by placing the particles at specific positions with respect to the illumination lasers. Again, monitoring the outgoing intensity at the fiber ends allows to continuously determine key properties of the collective motional states of the particles with minimal perturbation of the entanglement properties.

Although in our example we used a very simplified model system, it provides strong evidence that fiber coupled atoms could be a powerful, very versatile and scalable tool for quantum simulation as well as quantum computation. Certainly there are technical challenges to be met as e.g. ensuring sufficient radial confinement and cooling. In a more precise model one should also include back scattering inside the fiber and chirality related scattering control~\cite{iversen2021strongly, jones2020collectively,jen2021bound, mahmoodian2020dynamics,lodahl2017chiral}. This on the one hand could be very challenging to calculate but on the other hand a very promising extension towards even further control possibilities of this system.

\vspace{6pt} 

{\textbf{Contribution: }}{Conceptualization, D.H.; methodology, D.H., M.S. and H.R.; formal analysis, D.H.; investigation, D.H.; writing-original draft preparation, D.H. and M.S.; writing-review and editing, D.H., M.S. and H.R.; visualization, D.H.; supervision, M.S. and H.R.; project administration, H.R. All authors have read and agreed to the published version of the manuscript.}

{\textbf{Acknowledgements: }}{D.H. was funded by a DOC Fellowship of the Austrian Academy of Sciences ÖAW and H.R. acknowledges support from the FET Network Cryst3 funded by the
European Union (EU) via Horizon 2020.}

\section{Appendix}
\subsection{Data values for Figure~\ref{triangle1}}

Here we list the distances, frequencies and interaction strengths used to create Figure~\ref{triangle1} for different shapes of interacting ions.

\begin{table} 
\caption{Data values for Figure~\ref{triangle1} \label{data}}

\begin{tabular}{ccc}
 	&\textbf{Triangle}	& \textbf{Triangle with suppressed interactions}\\

\textbf{$\Omega_0/\widetilde{\Omega}$}	& 251.5&251.4\\
\textbf{$\Omega_1/\widetilde{\Omega}$}	& 643 & 642.6\\
\textbf{$\Omega_2/\widetilde{\Omega}$}	& 580.5&580.1\\
\textbf{$\Omega_3/\widetilde{\Omega}$}	& 72&72\\
\textbf{$\Omega_4/\widetilde{\Omega}$}	&0&0\\
\textbf{$\Omega_5/\widetilde{\Omega}$}	& 	666.2&665.8\\
\textbf{$\Omega_6/\widetilde{\Omega}$}	& 	1149.7&1149.1\\
\textbf{$\Omega_7/\widetilde{\Omega}$}	& 	754.3&754.3\\
\textbf{$\Omega_8/\widetilde{\Omega}$}	& 	104.7&104.8\\
\textbf{$\Omega_9/\widetilde{\Omega}$}	& 	115.5&115.3\\
\textbf{$\Omega_{10}/\widetilde{\Omega}$}	& 	591.3&590.8\\
\textbf{$\Omega_{11}/\widetilde{\Omega}$}	& 	724.8&724.5\\
\textbf{$\Omega_{12}/\widetilde{\Omega}$}	& 	392.4&392.4\\
\textbf{$\Omega_{13}/\widetilde{\Omega}$}	&   81.2	&81.3\\

\end{tabular}
\end{table}

\section{Time evolution of the coherent states}
\label{evolcoh}

In this section we go into more details on how the time evolution of coherent states $\vert\alpha\rangle\vert\alpha^\prime\rangle$
\begin{equation}
\hat{U}(t)\vert\alpha\rangle\vert\alpha^\prime\rangle=e^{-igt \left(\hat{a}_1\hat{a}_2^\dagger+\hat{a}_1^\dagger\hat{a}_2\right)}\vert\alpha\rangle\vert\alpha^\prime\rangle,
\end{equation}
with $\alpha,~\alpha^\prime$ taking values of $\pm \alpha$ can be calculated.

Using the definition of a coherent state, the facts that $\hat{U}(t)\vert 00\rangle=\vert 00\rangle$ and that $\hat{U}(t)$ is unitary, we can rewrite this equation
\begin{equation}
    \hat{U}(t)\vert\alpha\rangle\vert\alpha^\prime\rangle=e^{-\frac{\vert\alpha\vert^2+\vert\alpha^\prime\vert^2}{2}}\sum_{m,n=0}^\infty\frac{\alpha^m\alpha^{\prime n}}{\sqrt{m!n!}}\left(\hat{U}(t)\hat{a}_1^\dagger \hat{U}^\dagger(t)\right)^m \left(\hat{U}(t)\hat{a}_2^\dagger \hat{U}^\dagger(t)\right)^n\vert 00\rangle.
    \label{Ucoh}
\end{equation}
To evaluate $\hat{B}_i(t):=\hat{U}(t)\hat{a}_i^\dagger\hat{U}^\dagger(t)$ we use
\begin{equation}
    \frac{d}{dt}\hat{B}_i(t)=ig\hat{U}(t)\left[\hat{a}_1^\dagger\hat{a}_2+\hat{a}_1\hat{a}_2^\dagger,\hat{a}_i^\dagger \right]\hat{U}^\dagger(t)=ig\hat{A}_j,
\end{equation}
for $j\neq i$. The solution of this system of differential equations is
\begin{equation}
    \hat{A}_i=\hat{a}_i^\dagger\cos\left(gt\right)+i\hat{a}^\dagger_j\sin\left(gt\right).
\end{equation}
This way the sum in Equation~\eqref{Ucoh} can be rewritten to
   \begin{align}
   \sum_{m,n=0}^\infty\frac{\alpha^m\alpha^{\prime n}}{\sqrt{m!n!}} & \left(\hat{a}_1^\dagger\cos\left(gt\right)+i\hat{a}_2^\dagger\sin\left(gt\right)\right)^m \left(i\hat{a}_1^\dagger\sin\left(gt\right)+\hat{a}_2^\dagger\cos\left(gt\right)\right)^n\vert 00\rangle \nonumber \\
    &=e^{\alpha\left(\hat{a}_1^\dagger\cos\left(gt\right)+i\hat{a}_2^\dagger \sin\left(gt\right)\right)}e^{\alpha^\prime\left(i\hat{a}_1^\dagger \sin\left(gt\right)+\hat{a}_2^\dagger\cos\left(gt\right)\right)}\vert 00\rangle\nonumber\\
    &=e^{\hat{a}_1^\dagger\left(\alpha\cos\left(gt\right)+i\alpha^\prime\sin\left(gt\right)\right)}e^{\hat{a}_2^\dagger\left(\alpha^\prime i \sin\left(gt\right)+\alpha\cos\left(gt\right)\right)}\vert 00\rangle
\end{align} 
As $e^{-\alpha^\star\left(i\hat{a} \sin\left(gt\right)+\hat{a}\cos\left(gt\right)\right)}\vert 0\rangle=\vert 0\rangle$ and the displacement operator is defined as $\hat{D}\left(\alpha\right)=e^{-\vert\alpha\vert^2/2}e^{\alpha\hat{a}^\dagger}e^{-\alpha^\star\hat{a}}$, with $D\left(\alpha\right)\vert 0\rangle=\vert\alpha\rangle$ we find the desired result
\begin{align}
    \hat{U}(t)\vert\alpha\rangle\vert\alpha^\prime\rangle
    &=\hat{D}_1\left(\alpha\cos\left(gt\right)+i\alpha^\prime  \sin\left(gt\right)\right)\hat {D}_2\left( \alpha^\prime\cos\left(gt\right)+i\alpha\sin\left(gt\right)\right)\vert0\rangle\vert0\rangle \nonumber \\
   & =\vert \alpha\cos\left(gt\right)+i\alpha^\prime  \sin\left(gt\right)  \rangle\vert \alpha^\prime\cos\left(gt\right)+i\alpha\sin\left(gt\right)\rangle.
\end{align}


\clearpage


\clearpage
\section*{References}
\begin{thebibliography}{10}

\bibitem{kaufman2012cooling}
A.M. Kaufman, B.J. Lester, and C.A. Regal.
\newblock Cooling a single atom in an optical tweezer to its quantum ground
  state.
\newblock {\em Physical Review X}, 2(4):041014, 2012.

\bibitem{sheremet2021waveguide}
A.S. Sheremet, M.I. Petrov, I.V. Iorsh, A.V. Poshakinskiy,
  and A.N. Poddubny.
\newblock Waveguide quantum electrodynamics: collective radiance and
  photon-photon correlations.
\newblock {\em arXiv preprint arXiv:2103.06824v1}, 2021.

\bibitem{vetsch2010optical}
E.~Vetsch, D.~Reitz, G.~Sagu{\'e}, R.~Schmidt, S.T. Dawkins, and
  A.~Rauschenbeutel.
\newblock Optical interface created by laser-cooled atoms trapped in the
  evanescent field surrounding an optical nanofiber.
\newblock {\em Physical review letters}, 104(20):203603, 2010.

\bibitem{goban2012demonstration}
A.~Goban, K.S.~Choi, D.J.~Alton, D.~Ding, C.~Lacro{\^u}te, M.~Pototschnig, T.~Thiele,
  N.P.~Stern, and H.J.~Kimble.
\newblock Demonstration of a state-insensitive, compensated nanofiber trap.
\newblock {\em Physical Review Letters}, 109(3):033603, 2012.

\bibitem{beguin2018observation}
J.B. B{\'e}guin, J.H. M{\"u}ller, J.~Appel, and E.S. Polzik.
\newblock Observation of quantum spin noise in a 1d light-atoms quantum
  interface.
\newblock {\em Physical Review X}, 8(3):031010, 2018.

\bibitem{markussen2020measurement}
S.B. Markussen, J. Appel, C. {\O}stfeldt, J.B.S. B{\'e}guin, E.S. Polzik, and J.H. M{\"u}ller.
\newblock Measurement and simulation of atomic motion in nanoscale optical
  trapping potentials.
\newblock {\em Applied Physics B}, 126(4):1--5, 2020.

\bibitem{jones2020collectively}
R. Jones, G. Buonaiuto, B. Lang, I. Lesanovsky, and B. Olmos.
\newblock Collectively enhanced chiral photon emission from an atomic array
  near a nanofiber.
\newblock {\em Physical review letters}, 124(9):093601, 2020.

\bibitem{shomroni2014all}
I. Shomroni, S. Rosenblum, Y. Lovsky, O. Bechler, G. Guendelman,
  and B. Dayan.
\newblock All-optical routing of single photons by a one-atom switch controlled
  by a single photon.
\newblock {\em Science}, 345(6199):903--906, 2014.

\bibitem{pivovarov2021single}
V.A.~Pivovarov, L.V.~Gerasimov, J.~Berroir, T.~Ray, J.~Laurat, A.~Urvoy, and
  D.V.~Kupriyanov.
\newblock Single collective excitation of an atomic array trapped along a
  waveguide: a study of cooperative emission for different atomic chain
  configurations.
\newblock {\em arXiv preprint arXiv:2101.05398}, 2021.

\bibitem{holzmann2016tailored}
D. Holzmann and H. Ritsch.
\newblock Tailored long range forces on polarizable particles by collective
  scattering of broadband radiation.
\newblock {\em New Journal of Physics}, 18(10):103041, 2016.

\bibitem{prasad2020correlating}
A.S. Prasad, J. Hinney, S. Mahmoodian, K. Hammerer, S. Rind, P. Schneeweiss, A.S. S{\o}rensen, J. Volz, and A. Rauschenbeutel.
\newblock Correlating photons using the collective nonlinear response of atoms
  weakly coupled to an optical mode.
\newblock {\em Nature Photonics}, 14(12):719--722, 2020.

\bibitem{cirac2012atomic}
J.I. Cirac.
\newblock Atomic self-organization around tappered nanofibers.
\newblock In {\em Laser Science}, pages LW1J--6. Optical Society of America,
  2012.

\bibitem{chang2012cavity}
D.E. Chang, L.~Jiang, A.V. Gorshkov, and H.J. Kimble.
\newblock Cavity qed with atomic mirrors.
\newblock {\em New Journal of Physics}, 14(6):063003, 2012.

\bibitem{Metzger2006fiber}
N.K. Metzger, E.M. Wright, W.~Sibbett, and K.~Dholakia.
\newblock Visualization of optical binding of microparticles using a
  femtosecond fiber optical trap.
\newblock {\em Opt. Express}, 14(8):3677--3687, Apr 2006.

\bibitem{chang2013self}
D.E. Chang, J.I. Cirac, and H.J. Kimble.
\newblock Self-organization of atoms along a nanophotonic waveguide.
\newblock {\em Phys. Rev. Lett.}, 110:113606, 2013.

\bibitem{buonaiuto2021dynamical}
G. Buonaiuto, F. Carollo, B. Olmos, and I. Lesanovsky.
\newblock Dynamical phases and quantum correlations in an emitter-waveguide
  system with feedback.
\newblock {\em arXiv preprint arXiv:2102.02719}, 2021.

\bibitem{griesser2013light}
T. Grie{\ss}er and H. Ritsch.
\newblock Light-induced crystallization of cold atoms in a 1d optical trap.
\newblock {\em Physical review letters}, 111(5):055702, 2013.

\bibitem{holzmann2014self}
D. Holzmann, M. Sonnleitner, and H. Ritsch.
\newblock Self-ordering and collective dynamics of transversely illuminated
  point-scatterers in a 1d trap.
\newblock {\em Eur. Phys. J. D}, 68(11):352, 2014.

\bibitem{ostermann2014scattering}
S. Ostermann, M. Sonnleitner, and H. Ritsch.
\newblock Scattering approach to two-colour light forces and self-ordering of
  polarizable particles.
\newblock {\em New Journal of Physics}, 16(4):043017, 2014.

\bibitem{holzmann2018synthesizing}
D. Holzmann, M. Sonnleitner, and H. Ritsch.
\newblock Synthesizing variable particle interaction potentials via spectrally
  shaped spatially coherent illumination.
\newblock {\em New Journal of Physics}, 20(10):103009, 2018.

\bibitem{georgescu2014quantum}
I.M. Georgescu, S. Ashhab, and F. Nori.
\newblock Quantum simulation.
\newblock {\em Reviews of Modern Physics}, 86(1):153, 2014.

\bibitem{kim2021quantum}
E. Kim, X. Zhang, V.S. Ferreira, J. Banker, J.K. Iverson,
  A. Sipahigil, M. Bello, A. Gonz{\'a}lez-Tudela, M. Mirhosseini, and O. Painter.
\newblock Quantum electrodynamics in a topological waveguide.
\newblock {\em Physical Review X}, 11(1):011015, 2021.

\bibitem{feynman1982simulating}
R.P. Feynman.
\newblock Simulating physics with computers.
\newblock {\em Int. J. Theor. Phys}, 21(6/7), 1982.

\bibitem{hartmann2016quantum}
M.J. Hartmann.
\newblock Quantum simulation with interacting photons.
\newblock {\em Journal of Optics}, 18(10):104005, 2016.

\bibitem{longhi2011optical}
S. Longhi.
\newblock Optical realization of the two-site bose--hubbard model in waveguide
  lattices.
\newblock {\em Journal of Physics B: Atomic, Molecular and Optical Physics},
  44(5):051001, 2011.

\bibitem{noh2016quantum}
C. Noh and D.G. Angelakis.
\newblock Quantum simulations and many-body physics with light.
\newblock {\em Reports on Progress in Physics}, 80(1):016401, 2016.

\bibitem{tashima2019direct}
T. Tashima, H. Takashima, and S. Takeuchi.
\newblock Direct optical excitation of an nv center via a nanofiber
  bragg-cavity: a theoretical simulation.
\newblock {\em Optics express}, 27(19):27009--27016, 2019.

\bibitem{huo2012quantum}
M.X. Huo, C. Noh, B.M.~Rodr{\'\i}guez-Lara, and D.G. Angelakis.
\newblock Quantum simulation of cooper pairing with photons.
\newblock {\em Physical Review A}, 86(4):043840, 2012.

\bibitem{angelakis2013mimicking}
D.G. Angelakis, M.X. Huo, D. Chang, L.C. Kwek, and
  V. Korepin.
\newblock Mimicking interacting relativistic theories with stationary pulses of
  light.
\newblock {\em Physical review letters}, 110(10):100502, 2013.

\bibitem{davoudi2020towards}
Z. Davoudi, M. Hafezi, C. Monroe, G. Pagano, A. Seif, and A. Shaw.
\newblock Towards analog quantum simulations of lattice gauge theories with
  trapped ions.
\newblock {\em Physical Review Research}, 2(2):023015, 2020.

\bibitem{cirac1995quantum}
J.I. Cirac and P. Zoller.
\newblock Quantum computations with cold trapped ions.
\newblock {\em Physical review letters}, 74(20):4091, 1995.

\bibitem{kewes2016realistic}
G. Kewes, M. Schoengen, O. Neitzke, P. Lombardi, R.S.
  Sch{\"o}nfeld, G. Mazzamuto, A.W. Schell, J. Probst, J.
  Wolters, B. L{\"o}chel, et~al.
\newblock A realistic fabrication and design concept for quantum gates based on
  single emitters integrated in plasmonic-dielectric waveguide structures.
\newblock {\em Scientific Reports}, 6(1):1--10, 2016.

\bibitem{paulisch2016universal}
V. Paulisch, H.J.~Kimble, and A. Gonz{\'a}lez-Tudela.
\newblock Universal quantum computation in waveguide qed using decoherence free
  subspaces.
\newblock {\em New Journal of Physics}, 18(4):043041, 2016.

\bibitem{leong2020large}
W.S. Leong, M. Xin, Z. Chen, S. Chai, Y.~Wang, and S.Y.
  Lan.
\newblock Large array of schr{\"o}dinger cat states facilitated by an optical
  waveguide.
\newblock {\em Nature communications}, 11(1):1--7, 2020.

\bibitem{li2012robust}
Y. Li, L. Aolita, D.E. Chang, and L.C. Kwek.
\newblock Robust-fidelity atom-photon entangling gates in the weak-coupling
  regime.
\newblock {\em Physical review letters}, 109(16):160504, 2012.

\bibitem{gonzalez2011entanglement}
A.~Gonzalez-Tudela, D. Martin-Cano, E. Moreno, L. Martin-Moreno,
  C.~Tejedor, and F.J. Garcia-Vidal.
\newblock Entanglement of two qubits mediated by one-dimensional plasmonic
  waveguides.
\newblock {\em Physical review letters}, 106(2):020501, 2011.

\bibitem{snyder1983optical}
A.W. Snyder.
\newblock Optical waveguide theory.
\newblock {\em Springer US}, 1983.

\bibitem{le2005spontaneous}
F. Le~Kien, S. Dutta Gupta, V.~I. Balykin, and K.~Hakuta.
\newblock Spontaneous emission of a cesium atom near a nanofiber: Efficient
  coupling of light to guided modes.
\newblock {\em Phys. Rev. A}, 72:032509, Sep 2005.

\bibitem{scarpelli201999}
L. Scarpelli, B. Lang, F. Masia, D.M. Beggs, E.A. Muljarov, A.B. Young, R. Oulton,
  M~Kamp, S~H{\"o}fling, C~Schneider, et~al.
\newblock 99\% beta factor and directional coupling of quantum dots to fast
  light in photonic crystal waveguides determined by spectral imaging.
\newblock {\em Physical Review B}, 100(3):035311, 2019.

\bibitem{liu2018high}
F. Liu, A.J. Brash, J. O’Hara, L.M.P.P. Martins, C.L.
  Phillips, R.J. Coles, B. Royall, E. Clarke, C. Bentham,
  N. Prtljaga, et~al.
\newblock High purcell factor generation of indistinguishable on-chip single
  photons.
\newblock {\em Nature nanotechnology}, 13(9):835--840, 2018.

\bibitem{mirhosseini2019cavity}
M. Mirhosseini, E. Kim, X. Zhang, A. Sipahigil, P.B.
  Dieterle, A.J. Keller, A. Asenjo-Garcia, D.E. Chang, and O. Painter.
\newblock Cavity quantum electrodynamics with atom-like mirrors.
\newblock {\em Nature}, 569(7758):692--697, 2019.

\bibitem{wang2021high}
X. Wang, P. Zhang, G. Li, and T. Zhang.
\newblock High-efficiency coupling of single quantum emitters into
  hole-tailored nanofibers.
\newblock {\em Optics Express}, 29(7):11158--11168, 2021.

\bibitem{bogdanov2012quantum}
Y. I. Bogdanov, A.A. Kokin, V.F.
  Lukichev, A.A. Orlikovsky, I.A.
  Semenikhin, and A.Y. Chernyavskiy.
\newblock Quantum mechanics and development of information technology.
\newblock {\em Informatsionnye Tekhnologii i Vychslitel'nye Sistemy},
  1(1):17--31, 2012.

\bibitem{kok2007linear}
P. Kok, W.J. Munro, K. Nemoto, T.C. Ralph, J.P. Dowling,
  and G.J. Milburn.
\newblock Linear optical quantum computing with photonic qubits.
\newblock {\em Reviews of modern physics}, 79(1):135, 2007.

\bibitem{jeong2002efficient}
H. Jeong and M.S. Kim.
\newblock Efficient quantum computation using coherent states.
\newblock {\em Physical Review A}, 65(4):042305, 2002.

\bibitem{ralph2003quantum}
T.C. Ralph, A. Gilchrist, G.J. Milburn, W.J. Munro, and S.
  Glancy.
\newblock Quantum computation with optical coherent states.
\newblock {\em Physical Review A}, 68(4):042319, 2003.

\bibitem{marek2010elementary}
P. Marek and J. Fiur{\'a}{\v{s}}ek.
\newblock Elementary gates for quantum information with superposed coherent
  states.
\newblock {\em Physical Review A}, 82(1):014304, 2010.

\bibitem{iversen2021strongly}
O.A. Iversen and T. Pohl.
\newblock Strongly correlated states of light and repulsive photons in chiral
  chains of three-level quantum emitters.
\newblock {\em Physical Review Letters}, 126(8):083605, 2021.

\bibitem{jen2021bound}
H.H.~Jen.
\newblock Bound and subradiant multi-atom excitations in an atomic array with
  nonreciprocal couplings.
\newblock {\em arXiv preprint arXiv:2102.03757}, 2021.

\bibitem{mahmoodian2020dynamics}
S. Mahmoodian, G. Calaj{\'o}, D.E. Chang, K. Hammerer, and
  A.~S S{\o}rensen.
\newblock Dynamics of many-body photon bound states in chiral waveguide qed.
\newblock {\em Physical Review X}, 10(3):031011, 2020.

\bibitem{lodahl2017chiral}
P. Lodahl, S. Mahmoodian, S. Stobbe, A. Rauschenbeutel, P.
  Schneeweiss, J. Volz, H. Pichler, and P. Zoller.
\newblock Chiral quantum optics.
\newblock {\em Nature}, 541(7638):473--480, 2017.

\end{thebibliography}
\end{document}